\documentclass{article}
\usepackage[cp1251]{inputenc}
\usepackage[english,russian]{babel}
\usepackage[pdftex,unicode]{hyperref} %for pdftex
\usepackage{amsmath}
\usepackage{amsfonts}
\begin{document}
\begin{center}{\bf Spin states of para-water and ortho-water molecule in gas and liquid phases}

{\small V.K.Konyukhov}

{\it \footnotesize A. M. Prokhorov General Physics Institute, Russian Academy of Sciences, Vavilov Street 38, 119991 Moscow, Russia}\end{center}

{\small Spin degrees of freedom of water molecule in gas and liquid state were investigated in order to provide a reasonable answer about the unsolved problem of a long-term behavior of water spin isomers. The approach used involves an assumption that molecules change their spin state from a pure state to a mixed one when they interact with some sorts of adsorbent surface. Some models and conceptions of the quantum information processing were used.}\medskip

It is very interesting and quite surprising that spin isomers of water molecule show a long-term behavior of the ortho/para concentration in liquid water. It is worth noting that water spin isomers are accessible for investigation due to effect of selective adsorption of water molecules on solid surfaces. The first experimental observation of this effect has been performed in supersonic flow of rarefied gas. In a flowing gas that has consisted mainly of carbon dioxide molecules with a low admixture of water vapor solid nanoclasters were created due to carbon dioxide nucleation during the gas cooling. Water molecules were adsorbed on cluster's surface selectively of spin isomer because the adsorption probability for the para-molecules exceeded that of the ortho-molecules \cite{Kon_1986}.

To date three experiments are known that demonstrated the long-term behavior of ortho-water and para-water in liquid state for about one hour \cite{Kon_1995}, \cite{Tikh_2002} and about one week if water molecules dissolved in glycerin \cite{Makur_2007}.

The systems which will be considered in this article are the spin degrees of freedom of $\rm H_{2}\,^{16}O$ molecule in liquid water. It is assumed that water molecules do not have the rotational degree of freedom in liquid water, unlike the case of a rarefied gas. Each liquid water molecule is involved in about four hydrogen bonds, that preventing the free rotational moving. The breakage of these bonds usually leads to the rotation of the molecule around one of the remaining hydrogen bonds \cite{Chap_2009}. It is also assumed that molecules are in the ground vibrational and electronic states.

The starting point for the following consideration of the spin degrees of freedom is to be the state of water molecule in gas phase. Spin moment of two protons that belong to a nuclear frame of molecule create a linear superposition in the computational basis of the Hilbert space $\mathcal{H}=\mathbb{C}^{2}\otimes\mathbb{C}^2$ that can be decomposed into a direct sum of singlet and triplet states. The spin wave functions $\mid\psi\rangle$ of the two protons contains all basis vectors of $\mathcal{H}$ with equal coefficients

$$\mid\psi\rangle=\mid\Psi^{-}\rangle/2+(\mid\Psi^{+}\rangle+\mid00\rangle+\mid11\rangle)/2\qquad
\mid\Psi^{\pm}\rangle=(\mid01\rangle\pm\mid10\rangle)/\sqrt{2}$$

It is assumed that the spin degrees of freedom are well protected from interaction with the environment and one can regard this quantum system as a closed one. This important approach to the system state seems to be valid due to low energy of spin-rotational interaction and relatively rare molecule collisions.

The first term on the right-hand side of the equation for $\mid\psi\rangle$ is the spin state of the molecule associated with the para-water and the second one is  also the spin state associated with the ortho-water. The first term is constructed by only one basis vector while the second term consists of three basis vectors that results in the concentration of para/ortho=1/3 due to a different nuclear statistical weight of spin isomers. In other words, the number of the ortho-molecules is thrice as large as the number of the para-molecules in water vapor at the normal conditions. This is a well-known state of water molecules in the Earth atmosphere.

The second step in the consideration of the spin state of water molecules in gas phase is to introduce a certain mixed spin state, which appears as a result of adsorbent-molecular interaction. Experiments showed that this surface interaction produces a change in the para/ortho ratio in both directions between the preference for either para-water or ortho-water. In order to use a simple statistical approach to the describing of the para/ortho=1/3 ratio one must be quite sure that the population is the same for all spin levels. From a general point of view an assumption that the population levels are identical for all spin states is right because the energy level splitting due to spin-spin interaction and Zeeman splitting in Earth magnetic field is less than kT and therefore the spin levels may be considered to be degenerate. The pure state of the spin system is robust and does not allow the coefficients at the wave function to deviate from their given values. A possible way to prepare a mixed spin state $\rho_{gas}$ of water molecule after its interaction with absorbent is to use the pure singlet and triplet states and introduce probabilities of each portion.

$$\rho_{gas}=p\mid\Psi^{-}\rangle\!\langle\Psi^{-}\mid+\frac{1-p}{3}(\mid\Psi^{+}\rangle+\mid00\rangle+\mid11\rangle)(\langle\Psi^{+}\mid+\langle00\mid+\langle11\mid);\hspace{0.2cm} 0\le p\le 1$$

The state $\rho_{gas}$ has a few important and useful properties. At first, the probability $p$ may be changed in a wide range of available values that a water vapor may contain only para-molecules or ortho-molecules of high purity. The interaction of a gas molecule with adsorbent results in the appearance of composed quantum system including the gas molecule and the one contained in a thin layer of liquid water on the surface. The spin system of the two molecules may reasonably be regarded as a pure entangled state. When one of the molecules will be returned to the gas phase and the composed system breaks down the spin state of the free molecule must be only mixed due to the general rule of decay of no product bipartite quantum system \cite{Blum_1981}.

The free molecule has a rotational degree of freedom, which may be considered as environment acting upon the spin system. From a useable point of view, in the context of this article it is more   convenient to speak about all molecules that have interacted with adsorbent and have revealed the mixed structure of their spin state as spin-modified (SM) water molecules.
The rotational movement of spin-modified molecule acts  due to the spin-rotational coupling on  the spin system to mountain the mixed spin constriction that contains singlet and triplet terms via an effect of environment-induced superselection \cite{Schl_2004}. The spin-modified water shows a good long-term behavior of ortho/para concentration for low density gas \cite{Kon_1990}.

The last step is to find the spin state of spin-modified water molecule in liquid. 
It is now impossible to describe, in terms of transformations associated with a gas-liquid transition, how a certain SM-molecule from gas converts into one in liquid environment but one can define a few properties of the molecule that need to be retained during the transition.
It is reasonable to regard the SM-molecular transition from gas into liquid as one has been transmitted through a noisy quantum channel, therefore, the singlet part has been partly delivered without any change $(p^{\prime}<p)$ but the triplet part has been converted in a completely depolarized state \cite{Bennett 1995}.

Using the above assumptions one can write a new mixed state of the SM-molecule as

$$\begin{array}{c}
\rho_{liq}=p^{\prime}\mid\Psi^{-}\rangle\!\langle\Psi^{-}\mid+(1-p^{\prime})\frac{\mathbb{I}}{4}\\{}\\
\mathbb{I}=\mid\Psi^{-}\rangle\!\langle\Psi^{-}\mid+\mid\Psi^{+}\rangle\!\langle\Psi^{+}\mid+\mid00\rangle\!\langle00\mid+\mid11\rangle\!\langle11\mid
\end{array}$$

\noindent the spin state $\rho_{liq}$ of two protons of water molecule is the well-known Werner state for two qubits \cite{Voll_2001}. Each of the two states of $\rho_{liq}$ that is the singlet one and the second part state which is proportional to identity operator have an important property in the context of the spin isomer problem that they are to be maximally protected from interaction with the environment.

The spin state $\rho_{liq}$ of two protons of water molecule is the well-known Werner state for two qubits \cite{Voll_2001}. Each of two states of $\rho_{liq}$ that is the singlet one and the second part state which is proportional to identity operator have an important property in the context of the spin isomer problem: they are to be maximally protected from interaction with the environment.

The singlet state is invariant under a random perturbation if the environment acts on spin states of each proton in symmetric or collective way that means that the spins change their states simultaneously \cite{Lid_1999}. This approach is to be fulfilled because one has before used the superposition of the basis states to define the singlet one.

The high degree of Werner state protecting from interaction with the environment results in a zero magnetic moment of  a spin-modified water molecule in liquid state. It is easy to verify that a mean value of a collective spin operator $S_z$ in Werner state is zero via calculation of trace norm. It means that spin-modified water molecules in liquid state do not contribute to NMR signal.

This article gives the reasonable and well-defined answer about the unsolved problem, what occurs  with water molecule in gas that interacted with certain adsorbent so as the para/ortho ratio deviates from its traditional value 1/3. The answer is that the water molecule changes its spin state from the pure state to a mixed one (see also \cite{Kon_2009} in this context).\medskip

\renewcommand{\refname}{}


\begin{thebibliography}{99}

\bibitem{Kon_1986}V. K. Konyukhov, V. I. Tikhonov, T. L. Tikhonova, and V. N. Faizulaev, Pis’ma Zh. Tekh. Fiz., {\bf 12}, No 23, P. 1438, (1986)

\bibitem{Kon_1995} V.K.Konyukhov, V.P.Logvinenko, and V.I.Tikhonov, Bulletin of the Lebedev Physics Institute, No.6, 1995, p. 31-34, Allerton Press Inc./New York.

\bibitem{Tikh_2002} V.Tikhonov and A.Volkov, Science, {\bf 296}, 2363 (2002)

\bibitem{Makur_2007} A.M.Makurenkov, V.G.Artemov, P.O.Kapralov, V.I.Tikhonov, and A.A.Volkov, Radiophysics and Quantum Electronics, Vol. 50, Nos. 10–11, 832, (2007)

\bibitem{Chap_2009} M.Chaplin, arXiv:0706.1355

\bibitem{Blum_1981} K.Blum, {\it Density matrix theory and applications}, Plenum Press, 1981.

\bibitem{Schl_2004} M.Schlosshauer, Reviews of Modern Physics, {\bf 76}, 1267, (2004)

\bibitem{Kon_1990} V.K.Konyukhov, V.I.Tikhonov, T.L.Tikhonova, Proceedings of Institute of General Physics, {\bf 12}, (1990), p.208-215, Nova Science Publishers, Inc., New York. 

\bibitem{Bennett 1995} C.H.Bennett, D.P.DiVincenzo, J.A.Smolin, W.K.Wootters, arXiv: quant-ph/ 9604024v2 (1996)

\bibitem{Voll_2001} K.G.H.Vollbrecht and R.F.Werner, arXiv:quant-ph/0010095v2 (2001)

\bibitem{Lid_1999} D.A.Lidar, D.Bacon and K.B.Whaley arXiv:quant-ph/9809081v2 (1999)

\bibitem{Kon_2009} V.K.Konyukhov, Bulletin of the Lebedev Physics Institute, 2009, Vol. 36, No. 4, pp. 123-125., Allerton Press, Inc./New York.

\end{thebibliography}
\end{document}